\documentclass[fleqn,usenatbib]{mnras}

\usepackage{newtxtext,newtxmath}

\usepackage[T1]{fontenc}
\usepackage{ae,aecompl}

\emergencystretch=\maxdimen
\usepackage{graphicx}	
\usepackage{amsmath}	
\usepackage{amssymb}	

\title[Multi-telescope observations of FRB 121102]{Simultaneous multi-telescope observations of FRB 121102}

\author[M. Caleb et al.]{M. Caleb,$^{1}$\thanks{E-mail: manisha.caleb@manchester.ac.uk}
B. W. Stappers$^{1}$,
T. D. Abbott$^{3}$,
E. D. Barr$^{2}$,
M. C. Bezuidenhout$^{1}$,
\newauthor
S. J. Buchner$^{3}$,
M. Burgay$^{4}$,
W. Chen$^{2}$,
I. Cognard$^{5, 6}$,
L. N. Driessen$^{1}$,
R. Fender$^{7}$,
\newauthor
G. H. Hilmarsson$^{2}$,
J. Hoang$^{8}$,
D. M. Horn$^{3}$,
F. Jankowski$^{1}$,
M. Kramer$^{2}$,
\newauthor
D. R. Lorimer$^{9}$,
M. Malenta$^{1}$,
V. Morello$^{1}$,
M. Pilia$^{4}$,
E. Platts$^{10}$,
A. Possenti$^{4,11}$,
\newauthor
K. M. Rajwade$^{1}$,
A. Ridolfi$^{4,2}$,
L. Rhodes$^{7,2}$,
S. Sanidas$^{1}$,
M. Serylak$^{3,12}$,
\newauthor
L. G. Spitler$^{2}$,
L. J. Townsend$^{13,14}$,
A. Weltman$^{10}$,
P. A. Woudt$^{13}$,
J. Wu$^{2}$
\\ \\  
$^{1}$Jodrell Bank Centre for Astrophysics, Department of Physics and Astronomy, The University of Manchester, Manchester, M13 9PL, UK\\
$^{2}$Max-Planck-Institut f\"ur Radioastronomie, Auf dem H\"ugel 69, D-53121 Bonn, Germany \\
$^{3}$South African Radio Astronomy Observatory, 2 Fir Street, Black River Park, Observatory 7925, South Africa \\ 
$^{4}$INAF - Osservatorio Astronomico di Cagliari, via della Scienza 5, I-09047 Selargius (CA), Italy \\ 
$^{5}$Laboratoire de Physique et Chimie de l'Environnement et de l'Espace, Universit\'e d'Orl\'eans/CNRS, F-45071 Orl\'eans Cedex 02, France \\ 
$^{6}$Station de radioastronomie de Nan\c{c}ay, Observatoire de Paris, CNRS/INSU, F-18330 Nan\c{c}ay, France \\
$^{7}$Astrophysics, Department of Physics, University of Oxford, Keble Road, Oxford OX1 3RH, UK \\ 
$^{8}$Instituto de Part\`iculas y Cosmolog\`ia (IPARCOS), Universidad Complutense de Madrid, Madrid, Spain \\ 
$^{9}$Department of Physics and Astronomy, West Virginia University, PO Box 6315, Morgantown, WV 26506, USA \\ 
$^{10}$High Energy Physics, Cosmology \& Astrophysics Theory (HEPCAT) group, Department of Mathematics and Applied Mathematics, \\ University of Cape Town, 7701 Rondebosch, Cape Town, South Africa, \\
$^{11}$Universit\`a di Cagliari, Dipartimento di Fisica, S.P. Monserrato-Sestu Km 0.700, I-09042 Monserrato (CA), Italy \\ 
$^{12}$Department of Physics and Astronomy, University of the Western Cape, Bellville, Cape Town 7535, South Africa \\ 
$^{13}$Department of Astronomy, University of Cape Town, Private Bag X3, Rondebosch 7701, South Africa \\ 
$^{14}$South African Astronomical Observatory, P.O Box 9, Observatory, 7935, Cape Town, South \\ 
}

\date{Accepted XXX. Received YYY; in original form ZZZ}

\pubyear{2020}

\begin{document}
\label{firstpage}
\pagerange{\pageref{firstpage}--\pageref{lastpage}}
\maketitle

\begin{abstract}
We present 11 detections of FRB 121102 in $\sim3$ hours of observations during its `active' period on the 10th of September 2019. The detections were made using the newly deployed MeerTRAP system and single pulse detection pipeline at the MeerKAT radio telescope in South Africa. Fortuitously, the Nan\c{c}ay radio telescope observations on this day overlapped with the last hour of MeerKAT observations and resulted in 4 simultaneous detections. The observations with MeerKAT's wide band receiver, which extends down to relatively low frequencies ($900-1670$ MHz usable L-band range), have allowed us to get a detailed look at the complex frequency structure, intensity variations and frequency-dependent sub-pulse drifting.  The drift rates we measure for the full-band and sub-banded data are consistent with those published between $600-6500$ MHz with a slope of $-0.147 \pm 0.014$ ms$^{-1}$. Two of the detected bursts exhibit fainter `precursors' separated from the brighter main pulse by $\sim 28$ ms and $\sim 34$ ms. A follow-up multi-telescope campaign on the 6th and 8th October 2019 to better understand these frequency drifts and structures over a wide and continuous band was undertaken. No detections resulted, indicating that the source was `inactive' over a broad frequency range during this time.  

\end{abstract}

\begin{keywords}
instrumentation: interferometers -- intergalactic
medium -- surveys -- radio continuum: transients -- methods: data analysis
\end{keywords}



\section{Introduction}
Fast radio bursts (FRBs) are characterised by millisecond-duration, $\sim$Jy-level, bright radio pulses appearing at random locations in the sky and originating at cosmological distances \citep[e.g.][]{SriHarsh, Prochaska, JPMacquart}. However, the nature of the sources producing these FRBs is not known. Their large dispersion measures (DMs) which are the integrated electron column densities along the lines-of-sight, are believed to be effective proxies for distance \citep[e.g.][]{Lorimer, Thornton}. The measured DM values significantly exceed the maximum Galactic contribution \citep[see][for details]{frbcat} from the Milky Way's interstellar medium (ISM) implying distances of several gigaparsecs. The combination of cosmological origin of these pulses, along with their estimated high radio luminosities and correspondingly large brightness temperatures is what makes FRBs compelling. 
Almost every radio telescope in the world is currently undertaking large-area surveys at radio frequencies ranging from 100 MHz \citep{Coenen, KarastergiouFRB, TingayFRB} up to tens of GHz \citep{nat_Shannon, Michilli, Gajjar} to discover, study and understand these bursts. Over a hundred FRBs have been discovered to date of which only a handful have been localised to host galaxies. Of this sample of localised FRBs, only two have been known to repeat \citep{nat_spitler, Benito}.

The FRBs discovered to date show a remarkable diversity of observed properties in terms of spectral and temporal variations, polarization properties and repeatability. Perhaps, the most striking difference is the repeatability. Hundreds of hours of telescope time have been spent on the follow-up observations of known FRBs \citep[e.g.][]{Rane}, yet only some of them have been observed to repeat \citep{nat_spitler, nat_Shannon, Andersen} for a given sensitivity limit. Presently, instrumental sensitivity and time spent following up a known FRB field to look for repeats are the two major reasons for the observed dichotomy \citep{Caleb_NA, Kumar} if all FRBs are indeed repeating sources.

The nature of FRBs, and that of the repeating FRBs in particular, may be revealed from their emission properties. Hence, understanding the flux density spectra is of great importance. The first repeating FRB 121102 \citep{Spitler2014} has been localised to a low-metallicity, low-mass dwarf galaxy at $z \sim 0.2$ \citep{SriHarsh}, and studied extensively across multiple wavelengths and frequencies. Until now, most successful observations of this FRB have been carried out above 1 GHz \citep{Gourdji}, with bursts detected at frequencies as high as 8 GHz \citep[e.g.][]{Gajjar}. More recently, a single burst from this FRB was reported in the upper half of the CHIME 400 - 800 MHz band \citep{Josephy}.  The CHIME radio telescope has since discovered several more repeating FRBs \citep{Andersen}, of which one has been localised to a star forming region in a nearby massive spiral galaxy \citep{Benito}. The significant differences observed in the properties of localised repeating FRBs, such as in the value of the Faraday rotation measure and the existence of a luminous persistent radio counterpart, suggest that repeating FRBs originate in diverse host galaxies and local environments.

The spectral behaviour of FRBs seen so far is very unusual. This is especially true for FRB 121102, where the bursts are dominated by patches of bright emission with varying spectral indices across the bands. Though Galactic pulsars and magnetars are also seen to exhibit complex structures in the time domain, FRB 121102's pulses are strikingly different due to their enormous energies. High-time resolution ($\sim 10 \, \upmu$s) observations of FRB 121102 with the Arecibo telescope in the $1.1-1.7$ GHz band show bursts with complex time-frequency structures and sub-bursts \citep{Hessels}. The sub-bursts are seen to drift towards lower frequencies at later times within the burst envelope by $\sim 200$ MHz ms$^{-1}$ in the $1.1-1.7$ GHz band. Such spectro-temporal structure has been observed in other repeating \citep{Andersen} and as-yet-non-repeating FRBs \citep[e.g.][]{FarahFRB}. This observed structure is probably a combination of the unknown emission mechanism and line-of-sight propagation effects. While the downward drifting structure is more common in repeaters than non-repeaters, several repeat bursts do indeed occur without significant drifting sub-structure. Therefore, the lack of drifting structure in a burst is presently insufficient to classify the FRB as a non-repeater.

Continued long term monitoring of FRB 121102 is ongoing at various facilities around the world, to understand its nature and burst properties. It is evident from the literature over the years and these campaigns, that FRB 121102 exhibits sudden periods of activity with no statistically significant periodicity determined, followed by long periods of quiescence \citep{Oppermann, Gourdji, GZhang}. The Five-hundred-metre Aperture Spherical Telescope (FAST) in China reported several tens of detections of FRB 121102 pulses using their L-band (1.1 -- 1.5 GHz) array of 19-beams, between 29th and 31st August inclusive \citep{Li2019}. Consequently, we carried out observations of FRB 121102 during this `active' period as part of a Director's Discretionary Time (DDT) proposal at the MeerKAT radio telescope. 

In this paper we present our observations and detections of FRB 121102 with MeerKAT, which has allowed us to get a detailed look at a possible periodicity in the pulses due to the presence of `precursor' bursts, the complex frequency structure, and also the sub-pulse frequency drifting of some bursts. In Section \ref{sec:obsandanalyses} we present the observational configuration of the MeerKAT telescope, and our transient detection pipeline. In Section \ref{sec:campaign}, we present the subsequent multi-telescope campaign organised to study possible FRB 121102 bursts over a wide and continuous frequency band to study pulse spectral and temporal evolution. The results from our detections/non-detections are presented and discussed in Section \ref{sec:R&D}, following which we present our conclusions in Section \ref{sec:conc}.

\section{The MeerKAT radio telescope: Observations and Data Analyses}
\label{sec:obsandanalyses}
\label{sec:MKT}

The South African Radio Astronomy Observatory (SARAO) Meer(more) Karoo Array Telescope \citep[MeerKAT;][]{Jonas, Camilo18, Mauch} is a radio interferometer consisting of 64, 13.96\,m dishes in the Karoo region in South Africa. The dishes are spread out over 8 km with 40 dishes in the inner $\sim$1 km core. In these observations, MeerKAT is operating at a centre frequency of 1284 MHz and a usable bandwidth of $\sim770$ MHz. The Meer(more) TRAnsients and Pulsars (MeerTRAP) project at the MeerKAT telescope undertakes fully commensal, high time resolution searches of the transient radio sky, simultaneously with all the other ongoing MeerKAT Large Survey Projects. Using the MeerTRAP backend (see Section \ref{sec:tuse} for details), MeerKAT observes simultaneously in two modes: incoherent and coherent. In the coherent mode, the data from the inner 40 dishes in the $\sim1$ km core of the MeerKAT array, are coherently combined to form a number of beams on sky with a sensitive Field-of-View (FoV). Typically, in our commissioning mode we form 396 beams on sky with a combined FoV of $\sim 0.2$ deg$^{2}$. In the incoherent mode the intensities of all 64 MeerKAT dishes are added to create a less sensitive but much wider FoV of $\sim$1 deg$^{2}$.

The first DDT observations were carried out on the 10th of September 2019 starting at 03:43 UT for a duration of 3 hours and 11 bursts were detected (see Section \ref{sec:detections} for details). Consequently, we organised a multi-telescope follow-up campaign on the 6th and 8th of October 2019 (see Section \ref{sec:campaign} for details). In both cases, the observations were performed with the MeerTRAP backend system, which utilises state-of-the-art Graphics Processing Units (GPUs) allowing real time searches of the tied-array beams formed on the sky. For these observations, the well constrained position of FRB 121102 allowed us to use a phased array of 60 MeerKAT dishes. MeerKAT uses an FX correlator \citep{Camilo18} and the MeerTRAP transient search pipeline (see Section \ref{sec:tuse} for details) uses the F-engine output stream for ingest. The F-engine is configured to channelize the 856 MHz wide band into 4096 channels with a native time resolution of 4.785 $\upmu$s. The geometric delays as well as those determined from observations of a strong calibrator J0408$-$6545 are applied to the data stream in the telescope F-engine, thus phasing up the array. The complex voltage data are channelized and then sent over the Central BeamFormer (CBF) network to the beamforming User Supplied Equipment (FBFUSE) that was designed and developed at the Max-Planck-Institut f{\"u}r Radioastronomie in Bonn. FBFUSE combines this data from the dishes into the requested number of total intensity tied-array beams which are placed at the desired locations within the primary beam of the array. FBFUSE also combines time samples, to give an effective sampling rate of approximately 306.24 $\upmu$s. The beams are then put back onto the network where they are captured by the Transient User Supplied Equipment (TUSE), a real-time transient detection instrument developed by the MeerTRAP\footnote{\url{https://www.meertrap.org/}} team at the University of Manchester.

\subsection{Transient detection instrument}
\label{sec:tuse}

TUSE consists of 67 Lenovo servers with one head node and 66 compute nodes. Each compute node contains two Intel Xeon -- CPU processors, each possessing 16 logical cores for computation, two Nvidia GeForce 1080 Ti Graphical processing units (GPUs) and 256 GB of DDR4 Random Access Memory (RAM) blocks. Each of the nodes is connected to a breakout switch via 10 GbE network interface cards (NIC) that are used to ingest data coming from FBFUSE. Data from FBFUSE are received over the network on the NICs as SPEAD2\footnote{\url{https://casper.ssl.berkeley.edu/wiki/SPEAD}} packets that are read by the data ingest code and written to POSIX shared memory ring buffers of 50 seconds duration. The data are arranged such that each compute node processes a number of coherent beams to be processed in real-time. Since the data from the beamformer come in a frequency-time format (i.e. frequency being the slowest axis), they are transposed to a time-frequency format on a per beam basis, that are required by the search code. The resulting filterbank data are saved in separate shared memory buffers corresponding to each beam. More details on TUSE will be presented in an upcoming paper (Stappers et al. in prep).

For this targeted observation, since the position of FRB 121102 is well-known to within a few milliarcseconds, we were able to run in a mode where only 2 nodes with one beam per node were processing the data in real time. Only extracted candidates were saved for further examination.

\subsection{Single pulse search pipeline}
\label{sec:AA}

The data for each beam are searched for bright bursts using the state-of-the-art, GPU-based single pulse search pipeline AstroAccelerate\footnote{\url{https://github.com/AstroAccelerateOrg/astro-accelerate}} \citep{DimoudiWesley, Karel3, Karel2, Dimoudi, Karel}. The real-time search was done by incoherently de-dispersing in the DM range 0--5118.4~pc cm$^{-3}$, divided into multiple sub-ranges with varying DM steps and time averaging factors.
We also searched up to a maximum boxcar width of 0.67 s.
For the region containing the DM of the FRB 121102, 380.68--771.88~pc~cm$^{-3}$, the DM step and the effective sampling time were 0.652~pc~cm$^{-3}$ and 612.8~$\upmu$s respectively.
This particular choice of parameters allowed us to process all the data in real time, thanks to strict optimisations applied in the AstroAccelerate algorithms. 

To reduce the number of detections due to Radio Frequency Interference (RFI), we applied a static frequency channel mask to the data before the de-dispersion and single-pulse search.
The RFI remained stable throughout the observations, meaning that our choice of static mask was sufficient. We masked $\sim50\%$ of the band in the real-time search. We did not use this mask in the offline analysis of the detections in Section \ref{sec:detections} and instead cleaned the data manually, thereby masking $\sim 30\%$ of the band.
Additionally, the data were cleaned using standard zero-DM excision \citep{Zerodm_Eatough} to remove any remaining broadband RFI that was infrequent enough not to be included in the mask.
The extracted candidate files contained raw filterbank data of the dispersed pulse and additional padding of 0.5~s at the start and at the end of the file.

\begin{figure*}
\centering
\includegraphics[width=7.0 in]{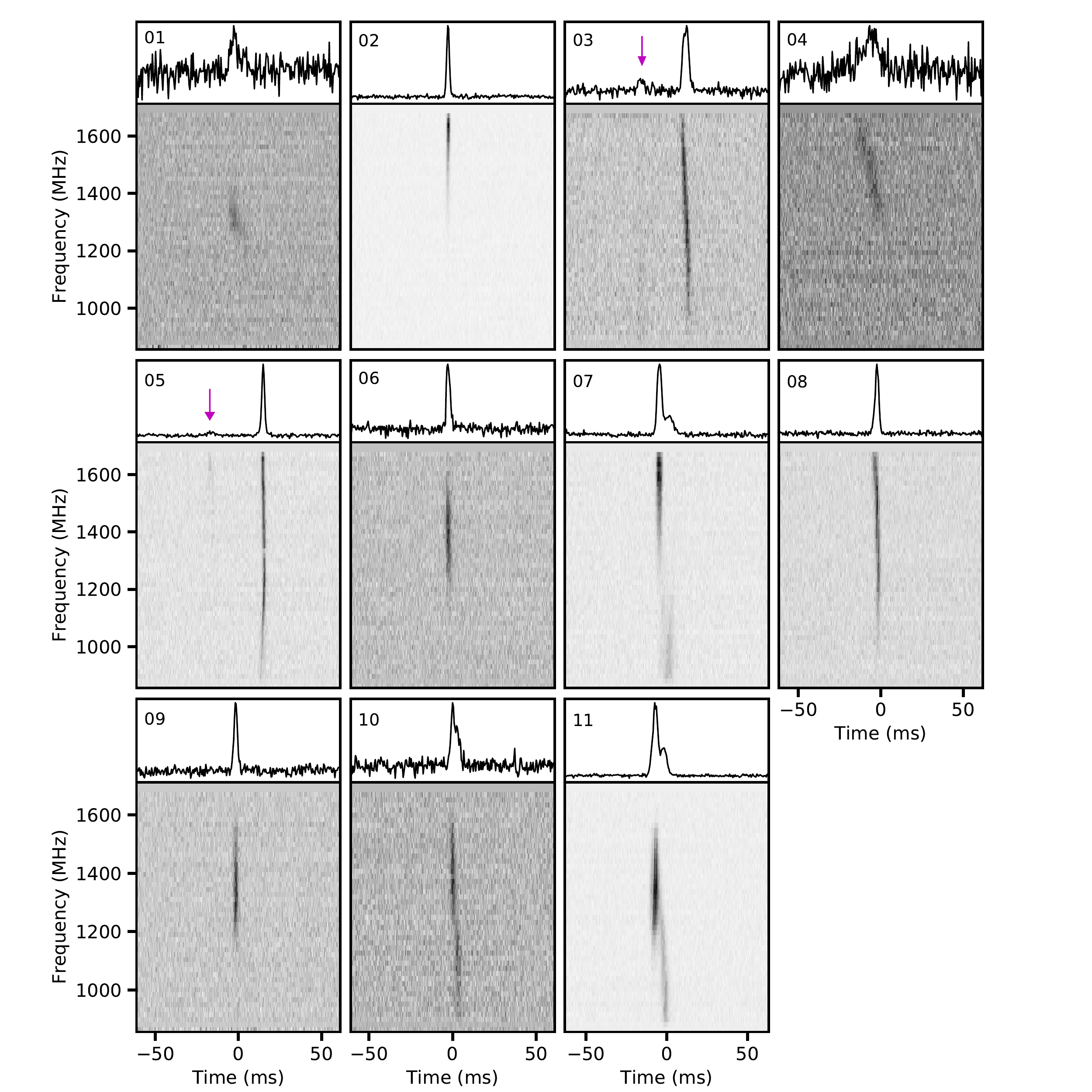}
\caption{Dynamic spectra of the bursts detected with MeerKAT. The top panel in each pulse shows the frequency-averaged pulse profile. The bottom panel shows de-dispersed frequency spectrum in which the frequency resolution of the bursts has been decimated to 64 channels each in order to make them more visible. The time resolution of the pulses is 306.24 $\upmu$s. Bursts 02, 06 and 09 have been de-dispersed to the S/N maximising DMs of 566.1, 563.2 and 565.2 pc cm$^{-3}$ respectively. All other bursts have been de-dispersed to an average DM of 564.8 pc cm$^{-3}$ (see Section \ref{sec:detections} for details). We note that we did not utilize the standard RFI excision mask detailed in Section \ref{sec:AA} and instead cleaned the data manually. The data are uncalibrated, and the flux densities are in arbitrary units. Bursts 03 and 05 are seen to show `precursor' bursts whose positions are indicated by the downward arrows.}
\label{fig:pulse_gallery} 
\end{figure*}

\begin{table*}
\caption{Observed properties of bursts detected with MeerKAT. We note that the S/N-optimised DM causes potential sub-bursts to overlap in time and sweep upward in frequency as seen in bursts 07 and 11 for example. As a result we do not quote DMs. Updated structure-optimized DMs and their errors will appear in Platts et al. (in prep). See Section \ref{sec:detections} for details on DM.}
\label{tab:properties}
\centering
\begin{tabular}{l c c c c c c c}
\hline\hline
Burst  & Arrival time & S/N & Width  & Fluence & Wait time \\ [0.5ex] 
            & (Topocentric MJD) &  & (ms) & (Jy ms) & (s)\\
\hline
01 & 58736.148545185752 & 12.2 & 6.3 & 0.18(2) &... \\
02 & 58736.165647624010 & 97.7 & 2.0 & 0.83(8) & 1477.65 \\
03 & 58736.165664566521 & 58.4 & 3.7 & 0.68(7) & 1.46 \\
04 & 58736.166942733849 & 15.5 & 12.0 & 0.32(4) & 110.43 \\
05 & 58736.184839564114 & 108.0 & 2.1 & 0.95(9) & 1546.28 \\
06 & 58736.200201955864 & 39.0 & 2.5 & 0.37(4) & 1327.31 \\ 
07 & 58736.211554436508 & 112.8 & 10.7 & 2.23(22) & 980.85 \\
08 & 58736.235176574322 & 88.7 & 2.6 & 0.87(9) & 2040.95 \\
09 & 58736.240623411504 & 45.6 & 2.3 & 0.42(4) & 470.60 \\
10 & 58736.247856676535 & 32.7 & 4.2 & 0.40(4) & 624.95 \\
11 & 58736.254215477995 & 208.8 & 10.7 & 4.13(41) & 549.40 \\
\hline
\end{tabular}
\end{table*}

\section{Multi-telescope campaign}
\label{sec:campaign}

Observing over the widest possible frequency range simultaneously would help us understand how the structure we see in the detections from our first DDT observations (see Figure \ref{fig:pulse_gallery}) relates to the drifting features seen in other publications, and also whether there are multiple instances of these apparent changes in burst properties as a function of frequency. Multi-frequency radio observations across a wide and continuous spectral range would provide valuable insight and new information on the emission/propagation effects. To this end, following the successful detections of 11 pulses from FRB 121102 on 10th September 2019, we organised a joint observing campaign on the 6th and 8th of October 2019, between the MeerKAT ($900 - 1670$\,MHz), Nan\c{c}ay ($1100 - 1800$\,GHz) and Effelsberg ($4000 - 8000$\,GHz) radio telescopes to better understand these frequency drifts and structures over a wide and continuous band.

\subsection{The MeerKAT radio telescope}

On October 6th and 8th 2019, the data were recorded according to the telescope specifications in Section \ref{sec:MKT}, and processed in real-time using the transient detection instrument and search pipeline detailed in Sections \ref{sec:tuse} and \ref{sec:AA}. In addition to the real-time processing in Section \ref{sec:tuse}, two additional nodes were used to record the data to disk at a full data rate for additional offline processing. No bursts were detected above a S/N of 10 by the real-time pipeline. The recorded data were processed offline for single pulses down to a lower S/N, as well as for a possible periodicity given the detections of `precursors' from the data taken on the 10th of September 2019. 

\subsubsection{Periodicity searches}
\label{sec:periodicity}

We incoherently de-dispersed the data recorded on the 6th and 8th of October 2019 over a range of trial DMs, $540.0 \leq \mathrm{DM} \leq 590.0$ pc cm$^{-3}$ in steps of 1 pc cm$^{-3}$. For each trial DM, the resulting de-dispersed time series was searched for short period pulsations using the Fast Fourier Transform (FFT). The FFT search was performed on the recorded data using the \textsc{Presto} suite of pulsar search and analysis software \citep{REM02}. The \texttt{realfft} routine was used to perform an FFT on each de-dispersed time series following which the \texttt{accelsearch} routine was used to sum 16 harmonics incoherently for each frequency bin to improve the S/N. All detections above a S/N of 8 were saved for further inspection. We folded the time series for each candidate period using the \texttt{prepfold} routine and the folded time series was visually inspected to check whether it resembled a true astrophysical source. No significant periodic pulsations were detected above the threshold S/N. For our threshold S/N, we report a flux limit of 0.2~mJy assuming a pulse duty cycle of 5$\%$.

A significant degree of sensitivity is lost during incoherent harmonic summing in FFT searches. Moreover, red-noise in the data can be an important factor when searching for longer periods. Therefore, we decided to perform a Fast Folding Algorithm search on the data using the \textsc{riptide}\footnote{\url{https://bitbucket.org/vmorello/riptide/src/master/}} FFA algorthm developed by Morello et al. (in prep) to search for periods ranging from 500 milliseconds to 10 minutes. The advantage of the FFA is that since we performed the search in the time domain, we did not lose any sensitivity to harmonic summing and were equally sensitive to a large range of pulse periods. Similar to the FFT, we de-dispersed and folded each time series and vetted the candidates for significant pulse profiles above a S/N of 8. We did not detect any significant periodic pulsations above the S/N threshold, which corresponds to a flux limit of 0.1~mJy for a 5$\%$ pulse duty cycle.

\subsubsection{Offline single pulse search}

We processed the recorded MeerTRAP data obtained on 2019 October 6 and 8 using the \textsc{heimdall} single-pulse search pipeline\footnote{\url{https://sourceforge.net/projects/heimdall-astro/}}. We performed both a coarse blind search and a finer targeted search centred around FRB~121102's nominal DM. The coarse search was done for trial DMs [0, 4000] pc cm$^{-3}$ with a 5~per cent S/N loss tolerance and the fine search was performed between DMs [0, 700] pc cm$^{-3}$ with a 1~per cent tolerance. The maximum boxcar filter width was 4096 samples, or about 313.6~ms. The frequency channels that were known to be affected by RFI at that time were masked during the \textsc{heimdall} run, leaving us with about 491~MHz of usable bandwidth. Standard zero-DM RFI excision was performed. We extracted the resulting candidates using the \textsc{dspsr} software package \citep{2011VanStraten} and we visualised them using \textsc{psrchive} \citep{2004Hotan} tools. We then visually inspected all candidates from both the coarse and fine search down to a S/N threshold of 6. No bursts were detected above the threshold of 0.04 Jy ms, assuming a 1 ms duration burst.

\begin{figure*}
\centering
\includegraphics[width=7.0 in]{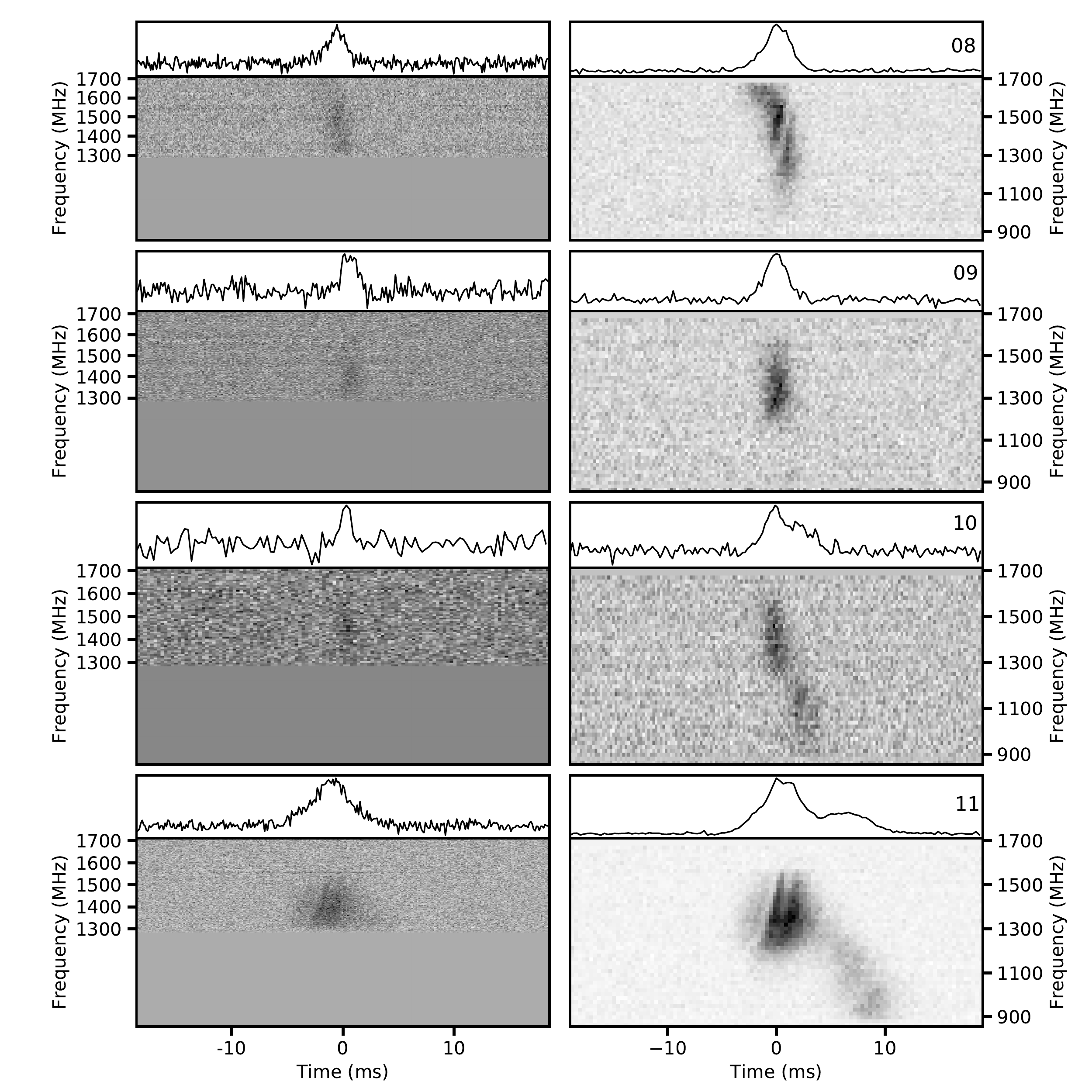}
\caption{Dynamic spectra of bursts 08, 09, 10 and 11 detected simultaneously with MeerKAT (856 MHz bandwidth) and Nan\c{c}ay (512 MHz bandwidth) radio telescopes. The top panel in each pulse shows the frequency-averaged pulse profile. The pulses detected with Nan\c{c}ay are plotted in the left column and the MeerKAT pulses are shown in the right column. The Nan\c{c}ay and corresponding MeerKAT pulses in each row have been de-dispersed to the DMs in Section \ref{sec:detections}. The four MeerKAT pulses have 64 frequency channels each, with a time resolution of 306.24 $\upmu$s. The Nan\c{c}ay pulses from top to bottom have 128, 64, 64 and 128 channels, with time resolutions of 256 $\upmu$s, 128 $\upmu$s, 1.024 ms and 128 $\upmu$s respectively.}
\label{fig:nancay_mkt_gallery} 
\end{figure*} 

\subsection{The Sardinia Radio Telescope}

The 64-m Sardinia Radio Telescope (SRT; \citealt{bolli+15}) observed FRB 121102 on the 6th of October starting at UT 01:45:42 for a total of 195 minutes. Baseband data were acquired with the LEAP ROACH1 backend \citep{bassa+16} covering a total bandwidth of 80 MHz centered at 336 MHz \citep[the P-band of SRT's coaxial L-P band receiver;][]{valente+10}. To minimise packet loss, the observation was split into 30-minute long segments (except for the last one, with a duration of 15 minutes) with $\sim 1$ minute gaps.

The search for single pulses was done offline using both the \textsc{Presto} and \textsc{heimdall} software packages. The first step of the analysis was to convert the baseband data, recorded in DADA\footnote{\url{http://psrdada.sourceforge.net/}} (Distributed Acquisition and Data Analysis) format into a filterbank format, compatible with both \textsc{Presto} and \textsc{heimdall}. This was done using the \texttt{digifil} routine of the \textsc{psrchive} package, creating a total-intensity 320-channels file, coherently de-dispersed at the dispersion measure of FRB 121102, with a sampling time of 4 $\upmu$s. The file was subsequently re-sampled with a final sampling time of 512 $\upmu$s. For the analysis based on \textsc{Presto}, we first removed RFI using \texttt{rfifind} (preliminarily flagging the frequency channels with the most prominent known RFI). The time series were de-dispersed with 200 DM steps covering the range 552-572 pc cm$^{-3}$ and analysed with \texttt{single$\_$pulse$\_$search.py} using a minimum S/N of 10. The high threshold was adopted because of the possibility of RFI severely affecting the P-band due to the absence of the shielding cover usually installed on the Gregorian dome during low-frequency observations at the SRT. The analysis with \textsc{heimdall}, also including a cross-match between candidates at adjacent DMs and times, was done on the same range of dispersion measures using a pulse width threshold of $2^5$ time bins of 16.384 $\upmu$s and a minimum S/N of 10. No reliable pulse candidates were found by either method above a fluence of 8 Jy ms assuming a 1 ms duration burst. Since the data were uncalibrated and the level of RFI severe, and given the uncertainty on the system temperature without Gregorian shielding, the stated fluence threshold is potentially a lower limit.

\subsection{The Effelsberg Radio telescope}
We observed FRB 121102 with the Effelsberg-100~m radio telescope
on 2019 October 6th and 8th for 240 and 360 minutes, respectively.
The observations started at 04:25:00 UTC on the 6th and
at 01:45:00 UTC on the 8th.
We used the S45mm single pixel receiver, which observes at 4--8~GHz
and has an SEFD of 25 Jy averaged across the band.
The data were recorded with full Stokes using two ROACH2 backends,
and were in a DADA format.
The data have 4096 channels, each with a bandwidth of 0.976562 MHz,
and a time resolution of 131 $\upmu$s. 

To search for single pulses, we used the \textsc{Presto}
software package.
In order to make our data compatible with \textsc{Presto} 
we extracted Stokes I from the data in a Sigproc filterbank format.
By using \texttt{rfifind} we created an RFI 
mask to apply to the data. We then created de-dispersed time series
between 0--1000 pc cm$^{-3}$ in steps of 2 pc cm$^{-3}$, in which
we searched for single pulses using \texttt{single$\_$pulse$\_$search.py}
with a S/N threshold of 7. No bursts were detected during our observations above a fluence of 0.06 Jy ms assuming a 1 ms duration burst.
The candidates were inspected by eye using \texttt{waterfaller.py},
which plots their dynamic spectra.

\begin{figure*}
\includegraphics[width=7.5 in]{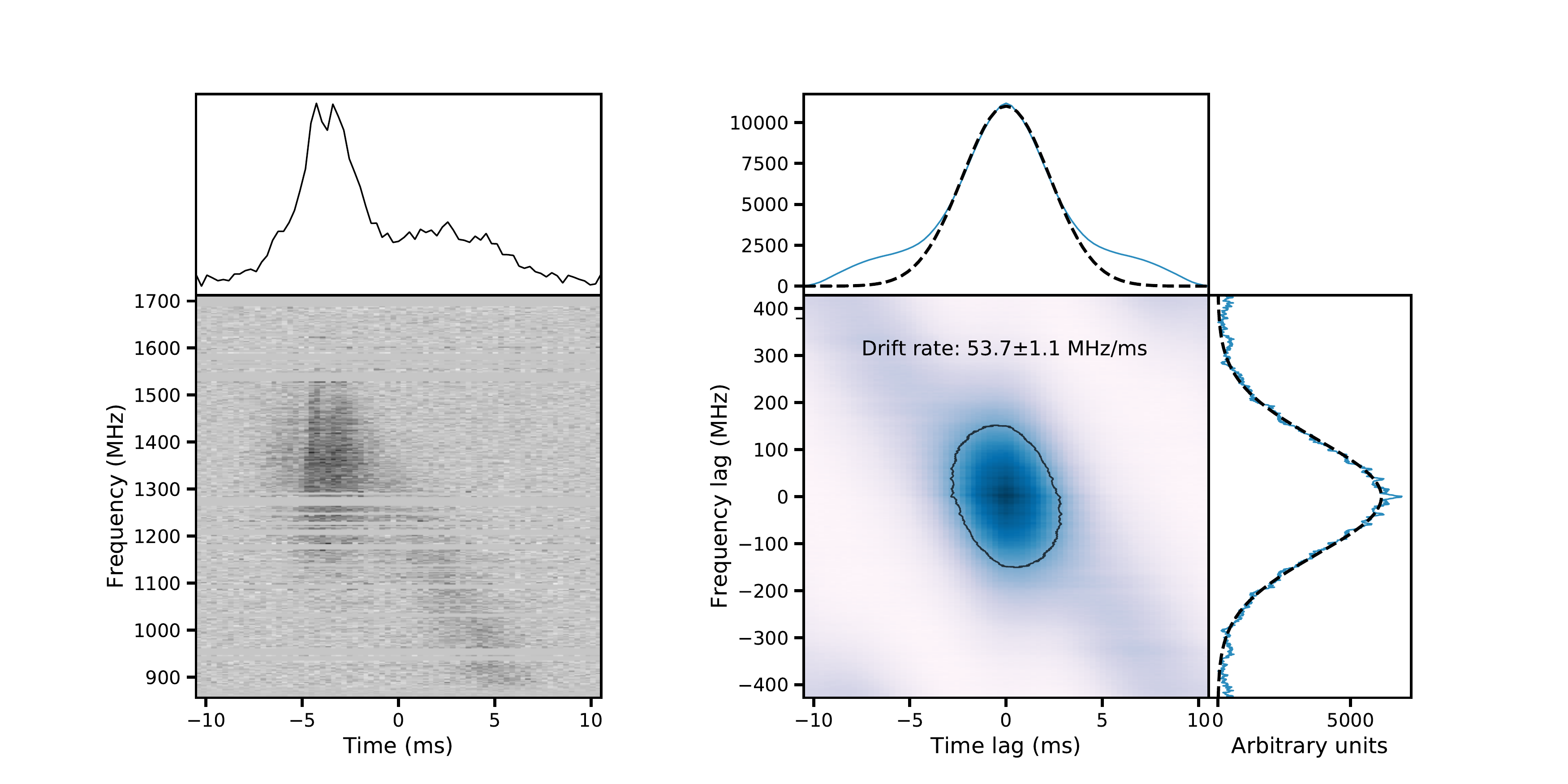}
\caption{\textit{Left}: Pulse profile of burst 11 with 1024 channels and 306.24 $\upmu$s time resolution, dedispersed to the structure maximising DM of 563.8 pc cm$^{-3}$. \textit{Right}: Autocorrelation function burst analysis for burst 11. The centre panel shows a two-dimensional ACF for the burst, with adjacent sub-panels showing the average along the time and frequency axes. These average ACF curves are fitted with Gaussian distributions.}
\label{fig:acf_drift} 
\end{figure*} 

\begin{figure}
\includegraphics[width=3.5 in]{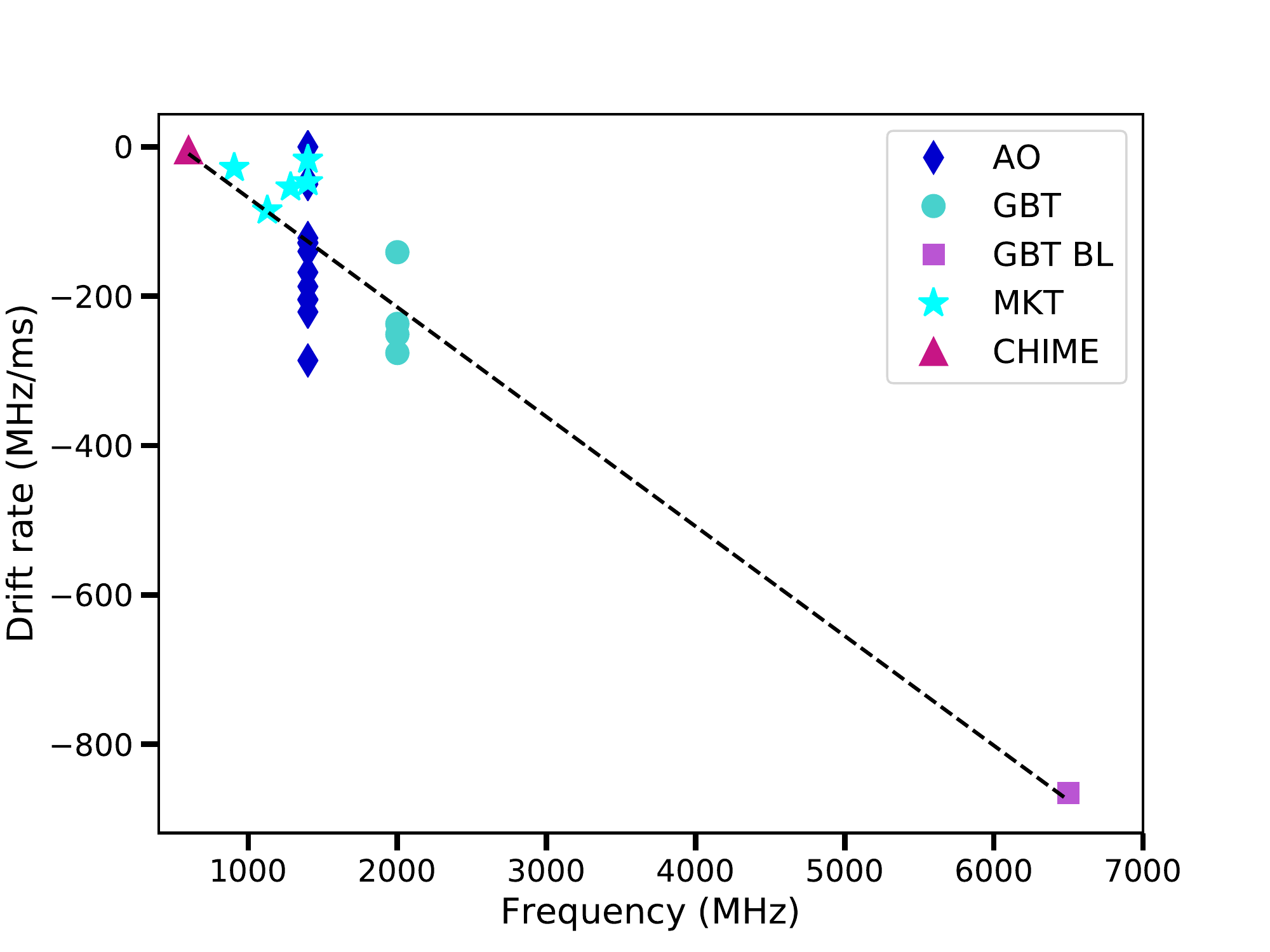}
\caption{Measured linear burst drift rate as a function of centre frequency for the ACF method. We fit a linear drift rate evolution with slope $\alpha = -0.147 \pm 0.014$ ms$^{-1}$.}
\label{fig:acf_driftrate} 
\end{figure} 

\subsection{Nan\c{c}ay Radio telescope}

The Nan\c{c}ay Radio Telescope (NRT) is a Kraus design meridian instrument equivalent to a 94-m dish. FRB 121102 was observed on the 10th of September and 6th of October 2019. On Sept 10th, the observation started at 05:29:40 UTC for 75 minutes while on Oct 6th, the observation started at 03:50:55UTC for 71 minutes.
Due to technical issue, no observation was performed on Oct 8th.
The Low Frequency receiver (1.1-1.8GHZ) was used and a total bandwidth of 512 MHz centred on 1486 MHz was processed through the search mode of NUPPI \citep[Nan\c{c}ay Ultimate Pulsar Processing Instrument;][]{Desvignes2011}. In this mode, 1024 channels of width 0.5 MHz are recorded as 4-bit in total intensity samples every 64 $\upmu$s.
The search for single dispersed bursts was performed offline using the \textsc{Presto} software. The data were cleaned with \texttt{rfifind} and de-dispersed with 128 DM values covering the range 527-591 pc cm$^{-3}$ with 0.5 pc cm$^{-3}$ steps.
The \textsc{Presto} script \texttt{single$\_$pulse$\_$search.py} searched for
individuals bursts using different averaging window widths.
All candidates above S/N of 6.5 were inspected by eye revealing 4 bursts shown in Figure \ref{fig:nancay_mkt_gallery} detected on the 10th of September, and none on the 6th of October above a fluence of 0.16 Jy ms assuming a 1 ms duration pulse.

\section{Results and Discussion}
\label{sec:R&D}

\subsection{MeerKAT detections}
\label{sec:detections}

The DDT observations on 10th September 2019 resulted in the detection of a total of 11 pulses from FRB 121102 in $\sim 3$ hours of observations. The pulse profiles are shown in Figure \ref{fig:pulse_gallery} and their properties are shown in Table \ref{tab:properties}. From Figure \ref{fig:pulse_gallery} it is evident that bursts 07, 10 and 11 exhibit sub-structure. The analysis of the sub-bursts and the structure-optimized DM will be reported in Platts et al. (in prep). We note that bursts 02, 06 and 09 are narrower in frequency and less likely to be affected by drifting substructure. As a result, bursts 02, 06 and 09 have been de-dispersed to the signal-to-noise (S/N) maximising DMs of 566.1, 563.2 and 565.2 pc cm$^{-3}$ respectively. For this work, all other bursts have been de-dispersed to the average DM of bursts 02, 06 and 09 (i.e. 564.8 pc cm$^{-3}$) in Figure \ref{fig:pulse_gallery}. The time series for each de-dispersed pulse was convolved with a series of Gaussian template profiles over a range of widths using the python based package \textsc{spyden}\footnote{\url{https://bitbucket.org/vmorello/spyden/src/master/}} to obtain the width and the best fit S/N for each pulse reported in Table \ref{tab:properties}. For reference, the intrachannel DM smearing time for a DM of 565 pc cm$^{-3}$ is $\sim 463 \, \upmu$s at the MeerKAT centre frequency of 1284 MHz. Therefore all the detected bursts are unresolved.

Incidentally, the Nan\c{c}ay radio telescope (NRT) was observing FRB 121102 on the same day and their observations overlapped with the last hour of MeerKAT observations. This resulted in the simultaneous detections of 4 bursts (Bursts 08, 09, 10 and 11 in Figure \ref{fig:nancay_mkt_gallery}), the analyses and details of which are discussed in Section \ref{sec:nancay_MKT}. Two of the MeerKAT detections (Bursts 03 and 05 in Figure \ref{fig:nancay_mkt_gallery}) show fainter `precursors' separated from the main pulse by $\sim28$ ms and $\sim34$ ms respectively with the signal level between the main burst and the precursor equal to the noise floor. We analysed our MeerKAT beam-formed data for the 11 pulses in great detail, and observed the spectra to vary across frequency as a function of time for most bursts. The MeerKAT detections reveal both narrow-band bursts and some that exhibit complex time-frequency structures and drifts. Similar spectral variation in bursts has also been reported by previous studies \citep{spitler, Michilli, Hessels} but never over such a large frequency range near 1.3 GHz. 

The sensitivity and large bandwidth of MeerKAT has resulted in a series of intriguing bursts which show structure and intensity variations across the entire band, as in Figure  \ref{fig:pulse_gallery}. 
We are able to probe down to the lower frequencies (rarely studied in the published bursts), and in most bursts there seems to be an interesting change at frequencies around 1200 MHz. At these frequencies, the source either becomes significantly fainter or we see complex bifurcated structure. It has been noted that in higher time resolution ($\sim 10 \, \upmu$s) observations, there is a tendency for the central frequency of a band-limited sub-burst to drift to lower frequencies at later times during the burst \citep{Hessels}. Though the downward drifting structure is more commonly observed in repeaters than non-repeaters, it should be noted that many repeat bursts do not have measurable frequency drift rates \citep{Michilli, Andersen}. It is also possible that the downward-drifting structure may be unresolved at coarse time resolutions.

\subsection{Simultaneous detections with MeerKAT and the NRT}
\label{sec:nancay_MKT}

Bursts 08, 09, 10 and 11 were detected simultaneously with the MeerKAT and the NRT on the 10th September of 2019 and are shown in Figure \ref{fig:nancay_mkt_gallery}. The burst structure observed in the 1200 - 1700 MHz band of MeerKAT is similar to the frequency characteristics we see in the corresponding NRT band. NRT's higher time resolution of 64 $\upmu$s compared to MeerTRAP's 306.24 $\upmu$s reveals sharp detailed structure in the high frequency half of the band, while MeerKAT's better sensitivity and larger bandwidth allow us to see more structure across the band. For reference, the intrachannel DM smearing times for 565 pc cm$^{−3}$ are $\sim 463 \, \upmu$s and $\sim 854 \, \upmu$s at the MeerKAT and NRT frequencies of 1284 MHz and 1400 MHz respectively.
Applying the radiometer equation with a system equivalent flux density (SEFD) of 25 Jy to flux calibrate the NRT data leads to peak fluxes of 0.61, 0.29, 0.17 and 1.06 Jy and widths of 3.2, 3.3, 4.0 and 5.1 ms respectively for the four detected bursts with the NRT. The estimated flux densities and measured widths are comparable to the MeerKAT values listed in Table \ref{tab:properties}. These simultaneous detections help us verify and characterise the MeerTRAP transient detection system.  

\begin{table}
\caption{Measured drift rates for various sub-bands of bursts 07 and 11.}
\label{tab:driftrates}
\centering
\begin{tabular}{l c c c c}
\hline\hline
Burst  & Centre frequency & Drift rate & Bandwidth \\ [0.5ex] 
       & (MHz) &  (MHz ms$^{-1}$) & (MHz) \\
\hline
07 & 1400 & -47.4$\pm$2.8 & 214 \\
\hline
11 & 906 & -27.5$\pm$0.5  & 100  \\
   & 1128 & -85.0$\pm$1.8 & 544 \\
   & 1284 & -53.7$\pm$1.1 & 856 \\
   & 1400 & -16.5$\pm$0.2 & 214 \\
\hline
\end{tabular}
\end{table}

\subsection{Drift rate analysis}
As evident from Figures \ref{fig:pulse_gallery} and \ref{fig:nancay_mkt_gallery} the MeerKAT pulses of FRB 121102 show intricate frequency structure similar to that reported by \cite{Hessels}. The analysis of the sub-bursts and the structure-optimized DM is not the focus of this paper and will be reported in Platts et al. (in prep).

The pulses in Figure \ref{fig:pulse_gallery} are seen to exhibit complex structure in frequency. \cite{Hessels} show that similar single peak pulses in their sample can be separated in distinct components called subbursts, when de-dispersed to the structure-optimized DM as opposed to the S/N-optimized DM. These subbursts are seen to exhibit a progressive drift towards lower frequencies at later times in the burst \citep[e.g.][]{Hessels}. Here we estimate the drift rates that were observed in select pulses from the sample of 11 bursts in Figure \ref{fig:pulse_gallery}. The method and analyses of the DMs which maximize structure in order to characterize sub-burst drifting will be presented in Platts et al. in prep. The structure optimized DM is more representative of the true DM of those bursts which exhibit structure. Not all pulses in the sample were observed to exhibit drifting structure. Though most appear to be sharp pulses, they could be a result of our relatively coarse time resolution of 306.24 $\upmu$s. However, some pulses from FRB 121102 indeed do not exhibit downward drifting sub-structure \citep{Michilli, Hessels}.

The on-pulse regions of the pulses after de-dispersing to the structure-optimized DM (see Platts et al. in prep) were extracted and a 2D Auto-Correlation Function (ACF) was computed. For each frequency channel, we compute the cross-correlation of the dedispersed signal as a function of time and frequency with a delayed copy of itself given by, 

\begin{equation}
    \mathrm{ACF}(\tau, \nu) = \int_{0}^{t} \int_{\nu}^{0} f(t,\nu) f(t-\tau, \nu-\nu') \, \mathrm{dt \, d\nu} ,
\end{equation}

\noindent where $\nu'$ and $\tau$ are the frequency and time lags respectively. An example of the ACF analysis for burst 11 is shown in Figure \ref{fig:acf_drift}. The zero-lag value, associated with self-noise, was excised from the ACF. Assuming  the  pulse  profile, spectral bandwidth of the burst envelope and subbursts are well described by a Gaussian, we measure their FWHMs from a Gaussian profile fit to the summed autocorrelation over the respective axes. A tilt in the autocorrelation ellipse reflects the drift rate in MHz ms$^{-1}$. Similar to \cite{Hessels}, we note that there is a tendency for the sub-bursts to drift to lower frequencies at later times during the burst. We successfully measured the drifts using the ACF method for bursts 07 and 11 (see Table \ref{tab:driftrates}). Given our large bandwidth, burst 07 was sub-banded and the drift rate was measured at a centre frequency of 1400 MHz over 214 MHz. The lower half of the structure optimised DM profile for burst 07 shows structure inconsistent with simple drifting (Platts et al. in prep) and is not used for this analysis. Burst 11 was similarly sub-banded and the drift rates were measured at four different centre frequencies over four different bandwidths (see Table \ref{tab:driftrates}). The drift rates at 1400 MHz and 1284 MHz were calculated using the entire available pulse (i.e. both components), while we only use the clearly sloped part of the profile (i.e. the second component), for the the measured rates at 1128 MHz and 906 MHz. We note that the 1400 MHz and 1284 MHz values for burst 11 are likely dominated by the bright non-drifting first component in Figure \ref{fig:acf_drift}. The measured drifts rates at these frequencies have allowed us to fill in the gap between 600 MHz and 1400 MHz in Figure \ref{fig:acf_driftrate}. Overall, the increase in magnitude of drift rate with increasing radio frequency is consistent with the results of \cite{Hessels}. We compare our measured drift rates with the ones published for this source between 600 - 6500 MHz in Figure \ref{fig:acf_driftrate} and fit a linear drift rate evolution with slope $\alpha = -0.147 \pm 0.014$ ms$^{-1}$, which is consistent with the measurement in \cite{Josephy}. We do not do a weighted fit as the uncertainties are not well quantified in the literature \citep{Hessels}. 

\subsection{Pulse periodicity}

Until recently, FRB 121102 was the only known repeating FRB and has generated much interest in searching for a possible periodicity in its pulses. The measurement of an emission period would be highly suggestive of progenitor models involving a rotating object. However, the lack of periodicity does not necessarily exclude rotating models. Two of the bursts detected by the Effelsberg radio telescope in 2017 were separated by $\sim 34$ ms \citep{Hardy}. Similarly, \cite{Scholz} report on two bursts detected by the Green Bank Telescope and separated by $\sim 37$ ms. Figure \ref{fig:pulse_gallery} presents our real-time detections of bursts 03 and 05 on the 10th September 2019 at UTs 03:58:33.419 and 04:26:10.138 respectively. Similar to the bursts detected by the Effelsberg telescope, each of our two bursts shows a small `precursor' separated from the main burst by $\sim28$ ms and $\sim34$ ms respectively with the signal level between the main burst and the precursor equal to the noise floor. It is apparent from Figure \ref{fig:pulse_gallery}, that the precursors in bursts 03 and 05 are fainter than the main pulse. The sensitivity and large bandwidth of MeerKAT have enabled us to detect these precursors, especially in the case of burst 05 in which the precursor spans a very narrow band. Interestingly, the precursor to burst 03 also drifts towards lower frequencies and looks similar to the brighter main pulse. Periodicity searches on recorded data from 8th October 2019 yielded no significant periodic pulsations above the threshold S/N (see Section \ref{sec:periodicity} for more details). The non-detection of a periodicity does not necessarily imply a limit on any possible underlying periodicity as the source could be a rotating object in which multiple bursts were emitted during a single rotation. Recent results suggest that though there is an observed periodicity in the activity of these repeaters, intrinsic periodicity between subsequent pulses is yet to be revealed \citep{ChimePeriodicity, Rajwade}.

\section{Conclusions}
\label{sec:conc}
In this paper, we present 11 detections of single pulses from FRB 121102 during its `active' period on 10th September 2019, using the newly commissioned MeerTRAP transient detection pipeline at the MeerKAT radio telescope in South Africa. Four of the 11 bursts (08, 09, 10 and 11) were detected simultaneously with the NRT thereby enabling us to verify and characterise our pipeline and system. Some pulses have sharp profiles while others exhibit downward drifting frequency sub-structure with increasing radio frequency. The analysis of the complex pulse structure and the structure-optimized DMs will be reported in Platts et al. (in prep). We characterize the drift rates of the pulses that exhibit sub-pulse drifting using a 2D ACF method. The measured drift rates at 906, 1284 and 1400 MHz are found to be consistent with those published between 600 - 6500 MHz, with a slope of $\alpha = -0.147 \pm 0.014$ ms$^{-1}$. Bursts 03 and 05 exhibit `precursors' separated from the main pulse peak by $\sim 28$ and $\sim 34$ ms respectively. 

A joint campaign of multi-frequency observations of FRB 121102 was carried out on 6th and 8th October 2019 between the MeerKAT (900--1670 MHz), Nan\c{c}ay (1100--1800 MHz) and Effelsberg (4000--8000 MHz) radio telescopes to better understand the frequency drifts and structures observed in the pulses detected in the previous observing run. The non-detections of bursts above the detection thresholds across this wide and continuous band in the radio meant that we could constrain the source to be in a truly `inactive' state during this period. Periodicity searches were performed on recorded data but no significant periodic pulsations were detected above the threshold. The lack of a detectable period does not exclude a rotating object as a progenitor, as the detections of precursors are suggestive of compact emission regions akin to a rotating object, in which multiple bursts are emitted during a single rotation.

\section*{Acknowledgements}
MC would like to thank SARAO for the approval of the DDT MeerKAT request and the CAM/CBF and operator teams for their time and effort invested in the observations. The MeerKAT telescope is operated by the South African Radio Astronomy Observatory (SARAO), which is a facility of the National Research Foundation, an agency of the Department of Science and Innovation. This work was partly based on observations with the 100-m telescope of the MPIfR (Max-Planck-Institut f{\"u}r Radioastronomie) at Effelsberg. The authors acknowledge funding from the European Research Council (ERC) under the European Union's Horizon 2020 research and innovation programme (grant agreement No 694745). LGS is a Lise Meitner independent research group leader and acknowledges support from the Max Planck Society. LR acknowledges the support given by the Science and Technology Facilities Council through an STFC studentship. 




\bibliographystyle{mnras}
\bibliography{refs} 







\bsp	
\label{lastpage}
\end{document}